# Antiferromagnetic Domain Wall Control via the Surface Spin Flop in Fully Tunable Synthetic Antiferromagnets with Perpendicular Magnetic Anisotropy

Benny Böhm[1], Lorenzo Fallarino[2], Darius Pohl[3,4], Bernd Rellinghaus[3,4], Kornelius Nielsch[3,4], Nikolai S. Kiselev[5] and Olav Hellwig[1,2]

[1]*Chemnitz University of Technology, 09107 Chemnitz, Germany*

[2]*Helmholtz-Zentrum Dresden-Rossendorf, Bautzner Landstraße 400, 01328 Dresden, Germany*

[3]*Dresden Center for Nanoanalysis, cfaed, TU Dresden, D-01062 Dresden, Germany*

[4]*IFW Dresden, Helmholtzstraße 20, 01069 Dresden, Germany*

[5]*Peter Grünberg Institut and Institute for Advanced Simulation, Forschungszentrum Jülich and JARA,*

*52425 Jülich, Germany*

## Abstract

Antiferromagnetic (AF) domain walls have recently attracted revived attention, not only in the emerging field of AF spintronics, but also more specifically for offering fast domain wall velocities and dynamic excitations up to the terahertz frequency regime. Here we introduce an approach to nucleate and stabilize an AF domain wall in a synthetic antiferromagnet (SAF). We present experimental and micromagnetic studies of the magnetization reversal in $[(Co/Pt)_{X-1}/Co/Ir]_{N-1}(Co/Pt)_X$ SAFs, where interface induced perpendicular magnetic anisotropy (PMA) and AF interlayer exchange coupling (IEC) are completely controlled via the individual layer thicknesses within the multilayer stack. By combining strong PMA with even stronger AF IEC, the SAF reveals a collective response to an external magnetic field applied normal to the surface, and we stabilize the characteristic surface spin flop (SSF) state for an even number N of AF-coupled $(Co/Pt)_{X-1}/Co$ multilayer blocks. In the SSF state our system provides a well-controlled and fully tunable vertical AF domain wall, easy to integrate as no single crystal substrates are required and with uniform 2D-magnetization in the film plane for further functionalization options, such as for example lateral patterning via lithography.

# Text

Antiferromagnets exhibit a wide range of interesting properties. Despite the fact that they do show zero net magnetization at remanence, antiferromagnets exhibit static and dynamic magnetic states, modes and types of reversal that reach beyond the functionality of ferromagnets [1, 2, 3, 4]. One classical example is the well-known bulk spin flop (BSF) state [5], where the components of the magnetic moments of the two antiferromagnetic (AF) sub-lattices form a canted state. With the development of synthetic antiferromagnets (SAFs) governed by interlayer exchange coupling (IEC) [6, 7] rather than the by much stronger direct atomic exchange interactions, AF characteristics could be successfully integrated into various applications [8, 9, 10, 11, 12]. In recent studies, SAF materials experience a scientific revival and become of interest in the field of modern spintronics [13, 14, 15, 16, 17]. SAF domain walls are for example promising candidates for high domain wall velocities [18] and SAFs can provide dynamic excitations that reach into the THz regime, similar to atomic antiferromagnets [19, 20, 21].

Theoretically predicted for semi-infinite layered antiferromagnets [22], the surface spin flop (SSF) occurs in SAFs where the AF-IEC overcomes a strong uniaxial anisotropy [23]. At the SSF-field, only those moments at the surface, which are aligned anti-parallel to the external field direction, rotate by about 90° with respect to the easy-axis. With increasing field strength, the canted SSF-state thus nucleates an AF domain wall, which is gradually propagated vertically down towards the center of the SAF [24]. From there, at higher external fields the canted BSF-like state in the center of the AF 180-degree domain wall expands over the whole layer stack, thus transforming it into the uniform BSF state [25]. The first experimental observation of the SSF was achieved by Wang et al. in 1994 [26] on epitaxially grown Fe/Cr SAFs on single crystal MgO(110) substrates with in-plane uniaxial magneto-crystalline anisotropy by comparison of superconducting quantum interference device (SQUID) and magneto optical Kerr effect (MOKE) magnetometry with an analytically derived model. Later, similar investigations were still limited to epitaxial systems with single crystal substrates [27, 28, 29].

In our study we use laterally isotropic $(Co/Pt)_{X-1}/Co$ multilayer (ML) blocks with strong interface induced perpendicular magnetic anisotropy (PMA), which are AF-coupled across 0.5 nm Ir spacer layers via very strong IEC [3, 30, 31] to form perpendicular magnetic anisotropy synthetic antiferromagnets (PMA-SAFs) of the type $[(Co/Pt)_{X-1}/Co/Ir]_{N-1}/(Co/Pt)_X$. The nomenclature is chosen such that N counts the total number of AF-coupled ferromagnetic (FM) ML blocks, while X refers to the number of Co layers per FM ML block. As shown in Fig. 1, for SAFs with *N>4* FM ML blocks with strong AF-IEC and strong PMA, two different collective reversal modes can be distinguished, depending on the dominating energy term, PMA or AF-IEC energy. Nevertheless, the remanent magnetic state in both cases corresponds to a laterally uniform AF-layered state (center image in Fig. 1) [3]. For dominating PMA energy, the reversal occurs via the formation of a complex 3-dimensional transitional state of mixed FM/AFM lateral stripe domains (see left hand side in Fig. 1) [3]. In this case, except for the lateral domain walls, all magnetic moments point along the easy anisotropy axis at all times of the magnetization reversal process. In contrast, for dominating AF-IEC energy, the field-induced magnetization reversal process occurs via a collective response by remaining in a laterally uniform state at any stage. Here the magnetization direction changes only normal to the film plane, thus representing a much simpler 1-dimensional heterogeneous state (see right hand side in Fig. 1). For even *N>4*, this reversal mode proceeds via the BSF state and the so called surface spin flop (SSF) state with a pinned AF domain wall forming normal to the film plane, where the close to 180° alignment of adjacent FM layers at any stage of the reversal reflects the dominance of the AF-IEC energy. For odd *N>4*, we observe an inversion (the SSF state is not stable here) from one into the other uncompensated uniform AF state via the formation of a normal-to-the-film-plane oriented AF domain wall, which for odd *N* can rapidly propagate vertically through the SAF without getting pinned. Note that the SAF system, we are using here, is characterized by a normal easy-axis along the natural ML growth direction and a 2-dimensional uniform hard anisotropy plane parallel to the layer structure. In contrast to previous epitaxially controlled SSF systems, our design leaves an additional degree of freedom normal to the easy-axis for further functionalization options for example via lithography.

The $\{[Co(0.5nm)/Pt(0.7nm)]_3/Co(0.5nm)/Ir(0.5nm)\}_{N-1}[Co(0.5nm)/Pt(0.7nm)]_4$ ML thin films were deposited via magnetron sputtering in a commercial UHV AJA ATC-2200 system. The depositions were performed at room temperature at an Ar working gas pressure of 3 mTorr with a Ta(1.5 nm)/Pt(20 nm) seed. A 2.3 nm Pt capping layer on top of the magnetic ML prevents the films from oxidation. Cross sectional (scanning) transmission electron microscopy (S)TEM was performed using an image and probe corrected FEI Titan[3] 80-300 operated at 300 kV accelerating voltage. Magnetic characterization was carried out using a Quantum Design SQUID vibrating sample magnetometer (VSM) at room temperature [32].

Exemplarily, cross sectional (S)TEM images of the *N=20* film are shown in Fig. 2. Fig. 2(a) displays a high-angle annular dark-field (HAADF) STEM image of the seed layer and ML structure. Columnar grains, which nucleate and develop during the growth of the Pt seed layer, become more pronounced within the SAF ML and are highlighted by dashed white lines. The 80 single Co(0.5 nm) layers are clearly visible. Additionally, a true-to-scale sketch of the nominal layer structure adjacent to the TEM image in Fig. 2(a) illustrates the good agreement between nominal and actual layer structures. In the TEM image of Fig. 2(b) also the larger, 4.6 nm thick [Co(0.5nm)/Pt(0.7nm)]₃/Co(0.5nm)/Ir(0.5nm) AF-coupled block structure with its 19 Ir coupling layers is visible. Overall, the columnar grain growth triggers the development of a correlated roughness, which increases with increasing ML thickness [33]. Fig. 2(c) shows a Fast-Fourier-Transform (FFT) of Fig. 2(a) and indicates the angular distributions of the elemental layering and the crystalline lattice planes. The elemental layering of Co, Pt and Ir shows a large angular variation of about 25° FWHM due to the developing correlated roughness. In contrast, the alignment of the out-of-plane atomic lattice planes remains about the same in the Pt seed as well as in the magnetic ML with a FWHM of about 7.5° (Fig. 2(c), outer lines).

Micromagnetic simulations were performed using the MuMax3 package [Vansteenkiste2014] for a qualitative macrospin model (Model I) and a more sophisticated quantitative multilayer model (Model II). Model I operates on a 32×32×$n$ grid, with one layer of cells representing a complete FM block, i.e. $n=N$ is the number of (Co/Pt)$_3$/Co blocks. Each cell has the dimension 3×3×4.1 nm$^3$. Due to the lateral homogeneity of the system, the external demagnetization fields are uniform and only contribute in the form of an easy-plane anisotropy, $K_d=\mu_0 M_s^2/2$. The saturation magnetization and effective uniaxial anisotropy were chosen to be $M_S$=775 kA/m and $K_{eff}=K_u-K_d$ to be 0.1 MJ/m$^3$, respectively, where $K_u$ is the interfacial anisotropy. The field is applied 0.57° tilted away from the z-axis in y-direction to mimic the non-perfect average alignment of anisotropy axis and external field axis due to the correlated ML roughness and crystallite angular distribution. The AF-IEC strength $J_{AF-IEC}$ was introduced by modifying the exchange stiffness $A$ (5×10$^{-12}$ J/m) between adjacent layers. After optimization of the parameters Model I shows qualitative agreement with the most significant features (BSF, SSF for even $N$ and inversion for odd $N$) for *N=20* and *N=21* with the same set of parameters. The corresponding MuMax script with optimized parameters is provided in the Supplementary Materials. The more sophisticated Model 2 captures each individual Co layer as a separate macrospin and includes a relative stronger AF-IEC for the Co layers in direct proximity to the Ir layers. A new parameter optimization with these additional aspects provides a more quantitative fit of the external field values and the net out-of-plane magnetization changes for BSF and SSF.

Fig. 3(a) shows the easy-axis out-of-plane magnetization curves of the *N=21* sample measured via SQUID-VSM and calculated by MuMax3. The inset illustrates the corresponding moment configurations as extracted from the micromagnetic simulations. Around remanence, two extended plateaus are observed with non-zero magnetization, originating from the uncompensated FM block. When increasing the external field towards saturation, the curves exhibit a single sharp transition at about ±1 T, followed by a curved segment that leads towards the uniform out-of-plane state. Moreover, at approx. ±0.5 T two sharp transitions through remanence are visible, switching back and forth between the two uncompensated AF-states with the top and bottom FM blocks being both down (i) or being both up in magnetization (ii) [34]. The two FM blocks at the surfaces only experience the AF-IEC on one side, while the bulk FM blocks are AF-coupled on both sides. Hence, the surface magnetic blocks start reversing at half the external field as compared to the bulk blocks. When one surface moment inverts in order to reduce its Zeeman energy, the strong AF-IEC will nucleate a vertical AF domain wall at the surface, which subsequently propagates vertically into the layer stack. If the other side of the SAF is not magnetically pinned by the Zeemann energy, the domain wall is fully propagated through the SAF with a full inversion of its AF configuration ((i) to (ii) or vice versa in Fig. 3(b)).

The large step at about ±1 T field is associated with the BSF transition from the AF state (ii) into the canted BSF state (iii), which lowers the Zeeman energy due to the equally directed z-components of all blocks along the external magnetic field direction. In this BSF state the AF order is still maintained, but limited to the hard-axis-component (within the x-y-plane). From the BSF state (iii) to saturation (iv) a gradual rotation of the moments into the field direction takes place. While theoretically for an ideal system the transition from the BSF state to saturation is linear in field with a sharp kink when reaching saturation (see simulation in Fig. 3(a)), we observe a more curved shape in the experiment, which proceeds towards saturation asymptotically. The more curved behavior is caused by either the correlated roughness (Fig. 2(a)), which leads to a laterally directional distribution in the anisotropy axes (especially at the grain boundaries), as well as slight deviations of the magnetic moment orientation of the Co layers within a single FM block along the *z*-axis, depending on the distance to the Ir interlayers. The strong AF-IEC directly at the Ir-interfaces may affect the Co moments more efficiently, thus making it harder to saturate them completely.

The magnetization curves and the corresponding simulated magnetic states of the *N=20* film are presented in Fig. 3(b). Clearly, a zero remanent magnetization is visible in the SQUID-VSM data and in the simulated curves in form of a large plateau region around remanence, reflecting the fully compensated magnetic moment for an even number of AF-coupled blocks. While applying a magnetic field, a subsequent two-step reversal behavior is observed. At about ±0.5 T an additional state occurs, as compared to the *N=21* sample. From the calculated magnetic states (inset of Fig. 3 (b)) the new additional state (ii) is identified as the SSF state. Here the flopped magnetization is localized only at the surface region, where the magnetic moment was previously aligned antiparallel to the external magnetic field direction. In contrast to the case *N=21*, here both surface moments in the AF state are antiparallel to each other. Hence, when rotating the surface moment pointing opposite to the applied field direction, the remaining layer stack exhibits a non-zero residual moment. Due to Zeeman energy minimization, the residual moment prevents the IEC from propagating the AF domain wall vertically through the entire SAF system (as previously observed for odd *N*). Instead, the angle of roughly 90° with respect to the easy-axis (center of the AF domain wall) moves to the center of the SAF stack (state (iii)). The small hysteresis at the SSF transition is associated with the nucleation vs annihilation of this vertical AF domain wall. The transition into the BSF state, (iii) to (iv) in Fig. 3 (b), occurs approximately at the same external field value as for *N=21* (compare the transition of state (ii) to (iii) in Fig. 3 (a)). In contrast to *N=21*, for *N=20* the BSF transition is not a vertically homogeneous flop within the whole layer stack. Instead, the already existing vertical AF domain wall with the locally flopped moments in the center of the SAF expands gradually out over the entire layer stack, until the system is fully transformed into the homogeneous BSF state (iv).

The quantitative discrepancies between the SQUID-VSM data and Model I are most prominent in the different magnetization step height, when transitioning into the BSF state and the subsequent qualitatively different behavior towards saturation. Those differences originate most likely from vertical sample asymmetries and lateral distributions with respect to the fitting parameters that are not captured in Model I, which only uses 20 or 21 all-identical macro spins with respect to magnetization, anisotropy and interlayer exchange. Thus, also any tilting of the spins within one ML block is forbidden. In order to get full qualitative agreement of Model I with the experiment for both samples with N=20 and 21, $J_{AF-IEC}$ was increased from 1.5 mJ/m² (as extracted from *N=2* SAF bilayer systems) to 3 mJ/m² per Co/Ir interface and the uniaxial PMA was reduced from 600

kJ/m$^3$ to 400 kJ/m$^3$. So, we had to over-emphasize the domination of the AF-IEC over the PMA in the Model I as compared to the experiment in order to get full qualitative agreement.

The more refined Model II is shown in Fig. 3(b) for *N=20*. Here, every [Co/Pt]$_3$/Co stack is represented by four individual magnetic layers, thus providing a more realistic description of the complex multilayer system. The magnetic parameters for Model II are optimized to $M_s$=800 kA/m, $K_{eff}$=600 kJ/m$^3$, $A$=5×10$^{-11}$J/m and $J_{AF-IEC}$=2 mJ/m$^2$. However, although this refined simulation shows much better quantitative agreement with the experiment, for the BSF transition height and the subsequent approach towards saturation for both samples, it is not able to fully capture the inversion for *N=21* around remanence. In contrast to the experimental observation during the magnetization reversal process in the simulations two domain walls nucleate simultaneously on both sides of the ML and the system gets frustrated. In our experiments, such a frustrated state is not observed, which we attribute to the increasing correlated roughness towards the top, thus breaking the top/bottom symmetry of the ML system. In Fig. 3(c) we display the various energy terms during magnetization reversal for the Model II with *N=20*. The domination of the AF-IEC energy over the PMA is clearly visible as the Zeeman energy drives the system through the reversal.

Using fully tunable double ML SAFs, easy to integrate and without the requirement of single crystal substrates, we demonstrated a pronounced SSF transition due to a dominating AF-IEC in the presence of a strong PMA. In contrast to previous experimental studies, our SSF system is designed with its easy anisotropy axis normal to the film plane, i.e. along the natural thin film direction of broken symmetry and thus provides for an additional in-the-plane (2-dimensional) homogeneous anisotropy landscape, which leaves room for subsequent nanostructuring and functionalization. The collective SSF response, as observed here for N=20, is transferable to all corresponding sample systems with even N>4, which we confirmed by observing a SSF state also for the lowest even N possible (i.e., for N=6) at otherwise identical sample parameters (not shown here). The SSF concept also works with synthetic ferrimagnetic systems, which will be useful to create a finite net magnetic moment for measuring dynamic modes of the pinned domain wall. Overall, we are confident that our findings and sample design open up a new level of AF domain wall control and tunability, which will be key for understanding AF-domain wall dynamics, interactions and the future role that AF domain walls in SAFs as well as atomic antiferromagnets can play in spintronic and magnonic applications.

# Acknowledgement

We would like to thank Eric E. Fullerton for fruitful discussions and Peter Matthies for help with the MOKE measurements.

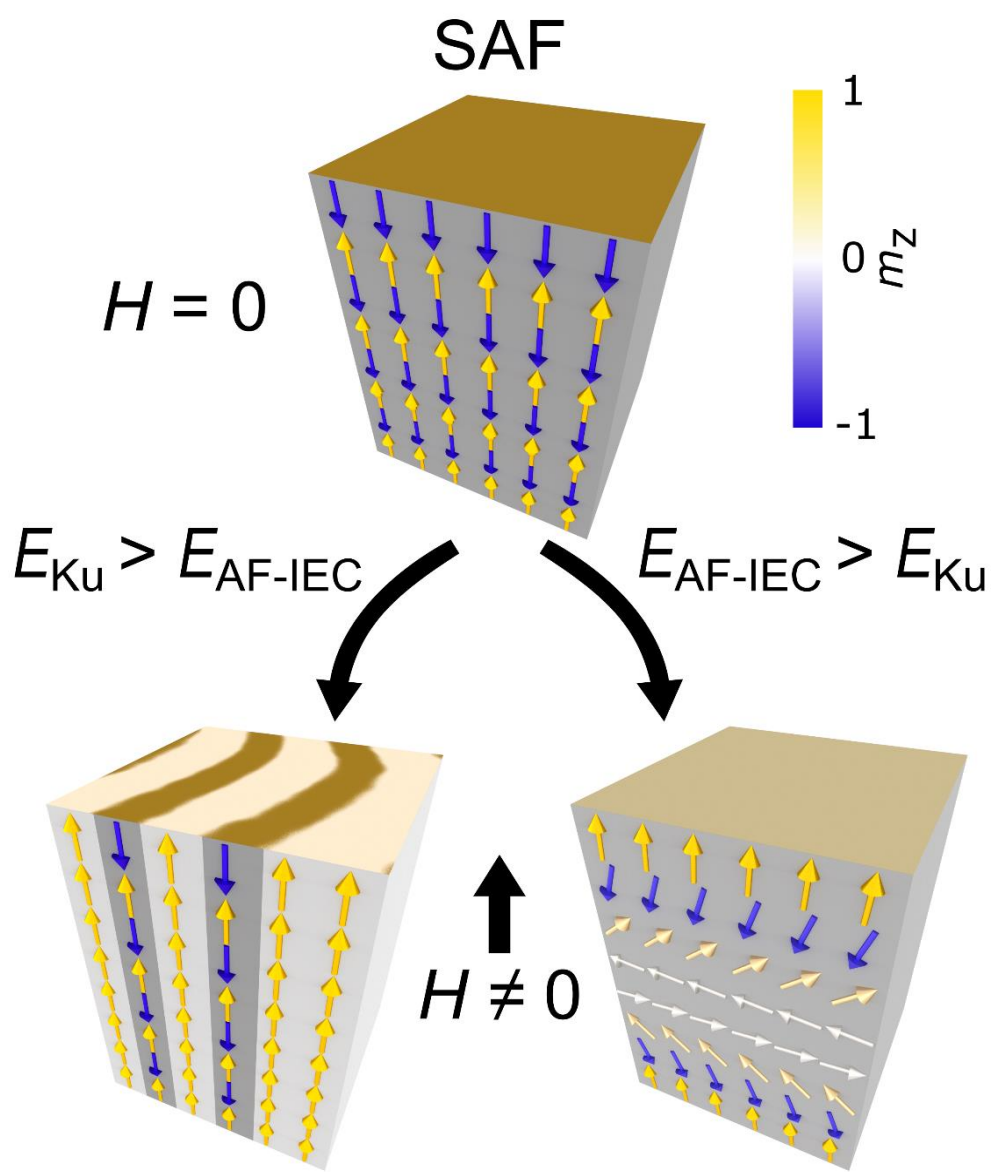

*Figure 1: Schematic collective reversal modes in SAFs, where PMA energy $E_{Ku}$ and AF-IEC energy $E_{AF-IEC}$ dominate over dipolar interactions. For $E_{Ku}$ > $E_{AF-IEC}$ the reversal proceeds via vertically and horizontally heterogeneous state of mixed FM/AFM stripe domains [Hellwig2007]. In contrast for $E_{AF-IEC}$ > $E_{Ku}$ the system shows heterogeneity during the reversal in only one direction normal to the film plane, while the system stays laterally homogeneous at all times (SSF state). Both reversal modes share the same laterally uniform AF layered remanent state. In our specific samples each arrow represents a FM (Co/Pt)$_3$/Co block and the coupling between adjacent blocks is mediated via 0.5 nm thick Ir spacer layers.*

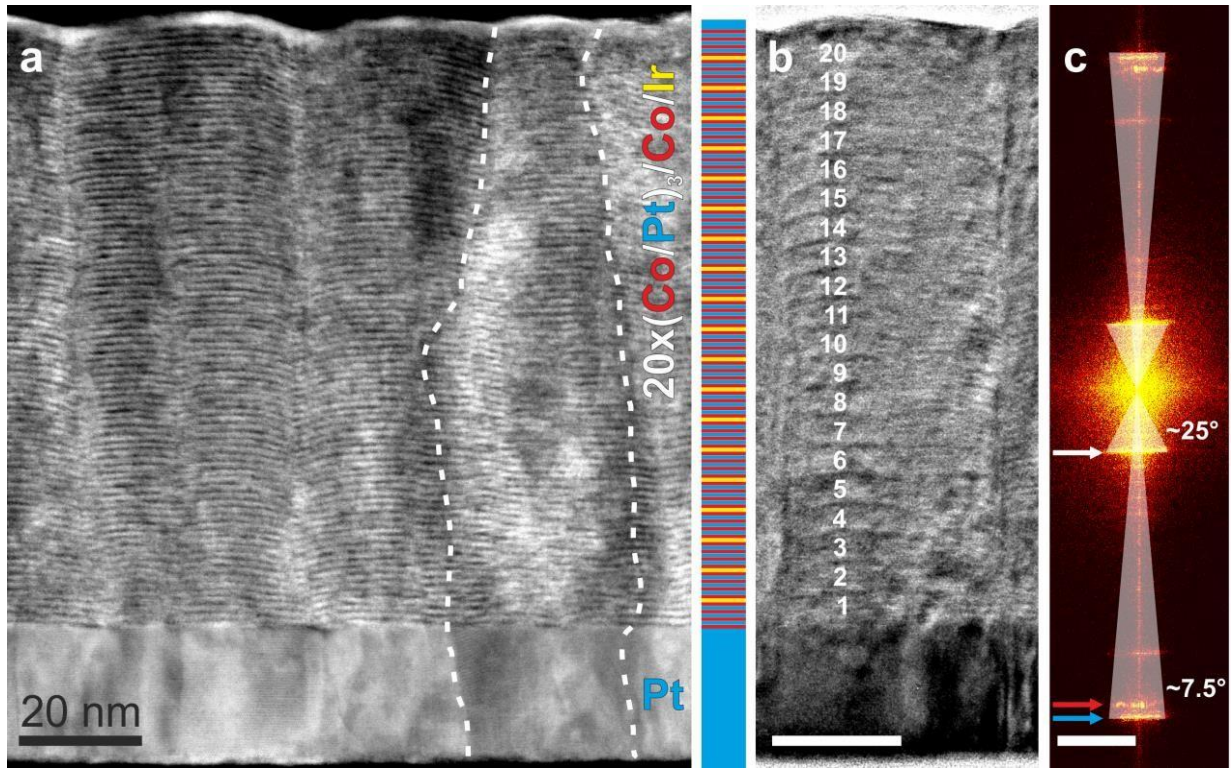

*Figure 2: Cross sectional TEM images of the N=20 SAF film. (a) High-resolution STEM image of the Pt seed and the SAF, where all 80 Co layers are resolved. The white dashed lines indicate that columnar grains nucleate in the Pt seed and then subsequently proceed through the entire SAF structure. A true-to-scale sketch illustration of the complete SAF ML structure is shown at the right (blue: Pt, yellow: Ir, red: Co). (b) HR-TEM image, where the [Co(0.5nm)/Pt(0.7nm)]$_3$/Co(0.5nm) - 4.1 nm thick block structure separated by Ir(0.5nm) layers becomes visible as illustrated by the white labeling of the 20 FM blocks. (c) FFT of (a) with scale bar 1/nm. The white cones indicate the angular distribution of the correlated ML roughness and the crystallite alignment. The arrows mark the Co/Pt ML period reflection (white) and the Bragg peak positions of the out-of-plane lattice spacing for the Pt-seed (red) and Co/Pt ML (blue) respectively.*

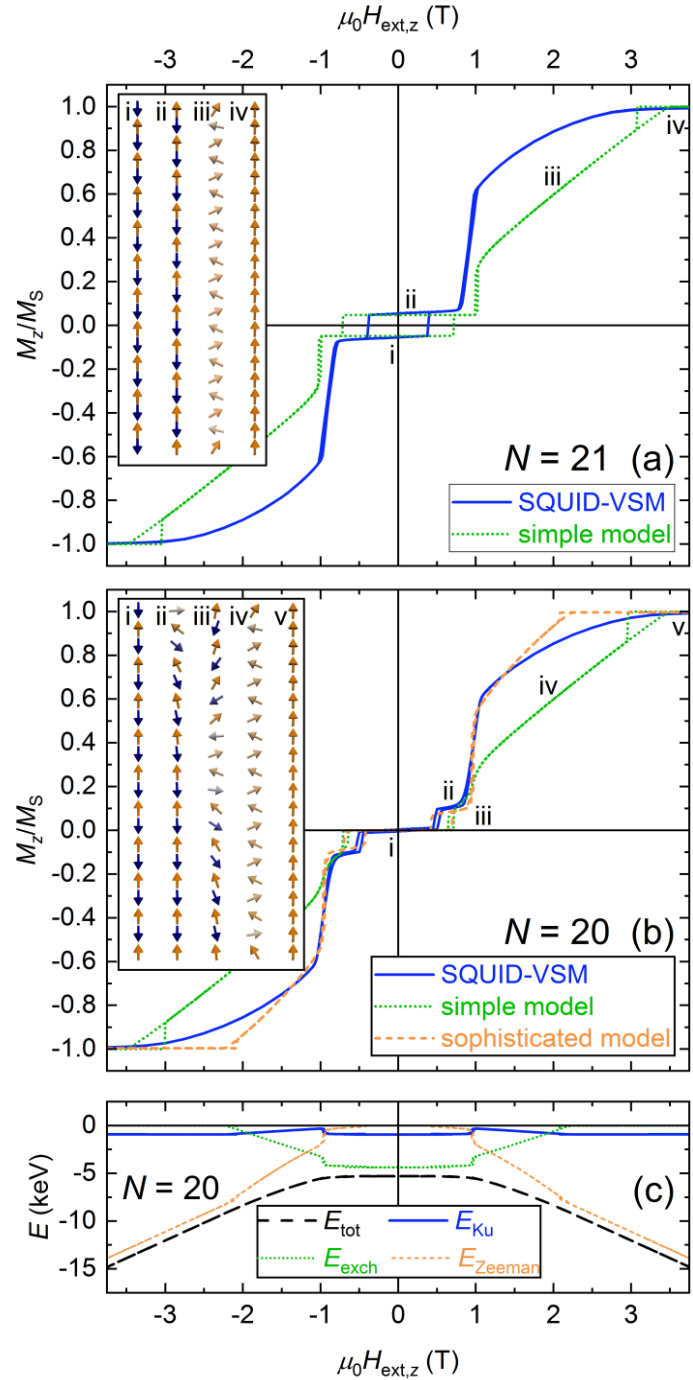

Figure 3: Out-of-plane hysteresis loops measured by SQUID-VSM and calculated by micromagnetic MuMax3 simulations for (a) N=21 and (b) N=20. The insets in (a) and (b) illustrate the corresponding magnetic configurations at the marked points (i-v) of Model I. (c) shows the energy balance of anisotropy ($E_{Ku}$), exchange ($E_{exch}$) and Zeeman ($E_{Zeeman}$) energy, which add up to the total magnetic energy ($E_{tot}$), for the refined Model II and N=20. The exchange energy contains the exchange stiffness A of the interatomic exchange as well as the IEC $J_{AF-IEC}$. The demagnetization energy is neglectable and set to zero here.